\documentclass{article}

\usepackage[english]{babel} 
\usepackage[utf8]{inputenc} 
\usepackage[T1]{fontenc} 
\usepackage{lmodern} 

\usepackage{graphicx} 
\usepackage{subcaption} 
\usepackage{multicol} 
\usepackage{float} 
\usepackage{wrapfig} 
\graphicspath{{./images/}}		

\usepackage[table]{xcolor} 
\usepackage{booktabs} 
\usepackage{multirow} 

\usepackage{amsmath} 
\usepackage{amsthm} 
\usepackage{amssymb} 

\usepackage{geometry} 
\usepackage{hyperref} 
\usepackage{xspace} 

\usepackage{algorithm} 

\newtheorem{theorem}{Theorem}[section]
\theoremstyle{definition}
\newtheorem{definition}[theorem]{Definition}

\usepackage{mathtools}

\usepackage{graphicx}

\usepackage{url}
\usepackage{amsmath}

\begin{document}

\title{Improving flocking behaviors in street networks with vision}

\author{Guillaume Moinard, Matthieu Latapy \\
Sorbonne Université, CNRS, LIP6, F-75005 Paris, France}

\date{}
\maketitle

\begin{abstract}
We improve a flocking model on street networks introduced in a previous paper. We expand the field of vision of walkers, making the model more realistic. Under such conditions, we obtain groups of walkers whose gathering times and robustness to break ups are better than previous results. We explain such improvements because the alignment rule with vision guaranties walkers do not split into divergent directions at intersections anymore, and because the attraction rule with vision gathers distant groups. This paves the way to a better understanding of events where walkers have collective decentralized goals, like protests.
\end{abstract}


\section{Introduction}
\label{sec:intro}

Consider the following scenario. Protesters are scattered throughout a city and share the common objective to gather into groups large enough to perform significant actions. They face forces that may break up groups, block some places or streets 
and seize any communication devices protesters may be carrying. As a consequence, protesters only have access to local information on
people and streets around them. Furthermore, formed protester groups must keep moving to avoid containment by adversary forces.

In this scenario, protesters need a distributed and as simple as possible protocol, that utilises local information exclusively and ensures a \emph{flocking} behavior, i.e., the rapid formation of significantly large, mobile, and robust groups.   

In a previous paper~\cite{moinard2024fast}, the city was modeled as a network of streets and intersections, and protesters were biased random walkers on this network. Authors then identified some key building block for those protocols to guaranty walkers will gather in a short range of time, such as the \emph{alignment} rule we will present in Section~\ref{sec:model}.

In this paper, while building on the same rules, we explore the effect of expanding walkers range of vision. Instead of applying their decision rules only to neighbor nodes in the graph, they will apply it to a set of nodes in a given range. We want to see if accessing more information in that manner improves walkers decision and enhances their flocking behavior.

In Section~\ref{sec:related}, we compare our approach to related works on flocking and gathering on networks. In Section~\ref{sec:framework}, we present the framework of street networks and walking rules introduced in~\cite{moinard2024fast}. We then define walkers vision in Section~\ref{sec:vision} and introduce new rules we construct with this vision. In Section~\ref{sec:experiments}, we run extensive experiments to measure the effect of vision on our walking rules and their combination. We also explore the robustness of groups, measuring how they reform if adversary forces break them up while following an effective tactic. Finally, we summarise our contributions in Section~\ref{sec:conclusion}.

\section{State of the Art}
\label{sec:related}

The most famous flocking model the Reynold's model~\cite{reynolds1987flocks}. Since its publication, many papers studied flocking behaviors in a wide variety of contexts.  However, most studies apply to continuous spaces, such as 2D or 3D spaces~\cite{vicsek1995novel,cucker2007emergent,olfati2007consensus,dorigo2021swarm}. In this paper, we focus on street networks, and such graphs are discrete spaces.

Nevertheless, some articles have explored algorithms for gathering on any connected graph~\cite{dessmark2006deterministic}. However, they require long term memory and computing capacities that exceeds what real pedestrians are capable of. Moreover, gathering is only a part of the flocking problem, as we also require walkers to subsequently move together once gathered.

Other articles have adapted rules proposed in the Reynold's model on a line of nodes~\cite{raymond2006flocking}. With respect to our article, this case corresponds to the situation of walkers in a single street. However, we want to have well-defined behaviors on the entire network, which includes the intersections.

In~\cite{moinard2024fast}, the authors presented a flocking model on street networks. Efficient behaviors emerged from a combination of multiple rules. We propose here a simpler and more realistic model inspired by Reynold's model, suited for any kind of network. Considering the importance of vision in the success of flocking models~\cite{giardina2008collective}, we introduce a vision range for walkers and study the impact of its depth on flocking.

\section{Framework}

\label{sec:framework}
We need a framework to simulate displacements of protesters in a city. We model cities 
as undirected graphs we call street networks. Protesters are then biased random walkers on this network. They can move from node to node with simple rules we introduce in the following Section.

\subsection{Discretized Street Networks}

In order to model real-world cities, we leverage OpenStreetMap data and the OSMnx library~\cite{boeing2017osmnx}. For a given city, we use this library with its default settings to extract the graph $G=(V,E)$ defined as follows: the nodes in $V$ represent street intersections in this city and the links in $E \subseteq V\times V$ represent pieces of streets between them. We take the undirected graph $G$, meaning there is no distinction between $(u,v)$ and $(v,u)$ in $E$. In addition, we denote by $N(v) = \{u, (u,v) \in E\}$ the set of neighbors of any node $v$ in $V$.

Like~\cite{moinard2024fast}, in the following, we use a typical instance, namely Paris, to present our work in this paper. This street network has 9\,602 nodes, 14\,974 links, leading to an average degree of 3.1. Its diameter is 83 hops and its average distance is 39.4 hops. The average street length is 99 meters, and the average distance is 5\,552 meters.

\begin{figure}
  \begin{center}
    \includegraphics[trim={0 0 0 5cm},clip,width=0.9\linewidth]{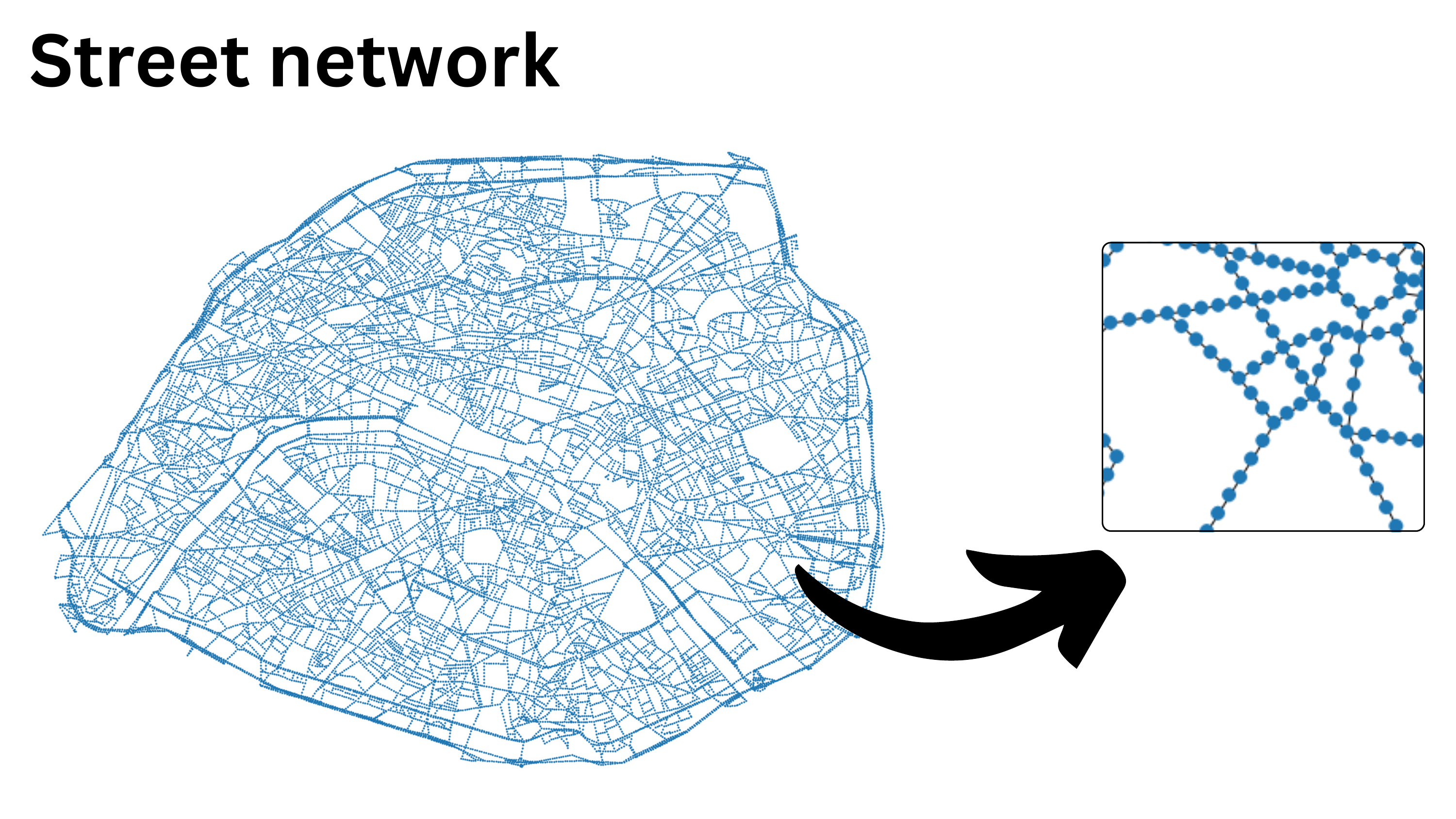}
    \caption{A piece of the discretized street network around {\em Place de la Nation} in Paris.}
    \label{fig:paris}
\end{center}
\end{figure}

The links of a street network generally represent street segments of very heterogeneous lengths~\cite{masucci2009random}. Then, moves from a node to another one may have very different duration. In order to model this, we use the same discretization procedure as in~\cite{moinard2024fast}. It consists in splitting each link of the street network into pieces connected by evenly spaced nodes. We illustrate this procedure with our Figure~\ref{fig:paris}.

In the obtained graph, each link represents a street slice of length close to a parameter $\delta$. Then a walker consistently make a move of length approximately $\delta$ at each hop.

Like in~\cite{moinard2024fast} we use $\delta$ equal to 10 meters, leading to a network of $N=130276$ nodes and $M=300736$ links. 

\subsection{Walkers}

\label{sec:model}

Given a network $G=(V,E)$, we consider a set $W$ of walkers numbered from $1$ to $|W|$. We denote the location of walker $i$ at time $t$ by $x_i(t) = v$, with $v\in V$. We call {\em group} the set of walkers at a given node $v$ at a given time $t$: $g_v(t) = \{i, x_i(t)=v \}$.  We denote by $g(t) = |\{g_v(t), v\in V, g_v(t)\ne \emptyset\}|$ the number of non-empty groups at step $t$.

We characterize \emph{flocking} as a gathering of walkers exploring the network. In~\cite{moinard2024fast}, big enough groups of walkers only form with rules that also make walkers explore the network. Therefore, the gathering of walkers is, on its own, a good indicator of the efficiency of our model.

\begin{definition}[Gathering score]
We denote the average size of non-empty groups of walkers by $\bar{n}(t) = \frac{W}{g(t)}$ and use this metric as our gathering score to evaluate the efficiency of our walks.
\end{definition}

At each time step $t$, a walker $i$ moves to a node $x_i(t+1) \in N(v)$. We present the two criteria, introduced in~\cite{moinard2024fast}, a walker uses to determine its next location:

\begin{enumerate}
    \item The \emph{number of walkers} located at node $u$ at time $t$; $n_u(t) = |g_u(t)|$
    \item The \emph{net flux} of walkers on a link $(u,v)\in E$ a walker perceives from node $u$; $J_{u\rightarrow v}(t)$ which is, at $t$, the difference between the number of walkers that entered the link from node $u$ at $t-1$, and those who entered from node $v$.
\end{enumerate}

Notice that the $J_{u\rightarrow v}(t)$ is a quantity that can be negative when the number of walkers that enter the link from $v$ is greater than those who enter from $u$. Moreover, we always have $J_{u\rightarrow v}(t) = -J_{v\rightarrow u}(t)$, which implies the flux can be an attractive or repulsive force for walkers, whether they stand at $u$ or $v$. We then can describe rules for the movement of walkers.

\begin{enumerate}
    \item \emph{Alignment rule:} A walker at node $u$ at time $t$ moves to the neighbor $v$ that maximizes the net flux $J_{u\rightarrow v}(t)$.
    \item \emph{Attraction rule:} A walker at node $u$ at time $t$ moves to the neighbor $v$ that maximizes the number of walkers $n_v(t)$.
\end{enumerate}

So far those rules do not take into account the vision of walkers. We now present the vision-based rules that consider the depth of the vision of walkers.

\section{Vision}
\label{sec:vision}

We construct walkers vision in the following manner. We introduce the concepts of \emph{vision depth} and \emph{branches} to define how a walker combines the rules from the previous section with a wider field of vision. 

\begin{definition}[Vision depth]
A walker can see what happens on the network up to a certain distance from its location, i.e., the vision depth $d$. This depth is an integer, the distance between two nodes being the number of links in a shortest path connecting them.
\end{definition}

A vision depth equal to $1$ means a walker $i$ can evaluate the number of walkers on its neighbor nodes $v\in N(x_i(t))$ and the net flux on adjacent links $(x_i(t),v)$. It sees all information at distance $1$ on the network. 
If this depth is greater than $1$, a walker perceives the two criteria on nodes and links further than its neighborhood. On the other hand, the special case of a vision equal to $0$ corresponds to a blind walker that can only follow a random walk.

Notice the depth of a walker vision can differ from one criterion to another. We call respectively $d_n$ and $d_j$ the depth of the vision for the number of walkers and for the net flux. A walker can see those quantities on nodes or links up to a distance $d_n$ and $d_j$ respectively.

Within this field of vision, a walker can look into different directions, at least one per neighbor nodes. We say a walker perceives multiple \emph{branches}.

\begin{definition}[Branch]
    First, a branch $B$ is a simple path, i.e., a sequence of nodes $(b_0, b_1, ..., b_k)$ such that $\forall i\neq j, b_i \neq b_j$ and $\forall i, (b_i, b_{i+1}) \in E$. Moreover, a branch must not be the prefix of any other simple path.  We denote by $\mathcal{B}(u)$ the set of all branches starting with $b_0=u$.
\end{definition}

A walker $i$ considers all branches starting from its location $x_i(t)$. This means that, $\forall B\in\mathcal{B}(x_i(t))$, it evaluates the values of the two criteria respectively up to the $d_n$ first nodes and $d_j$ first links of the branch $B$.

\begin{figure}[h!]
  \begin{center}
      \includegraphics[width=0.75\linewidth]{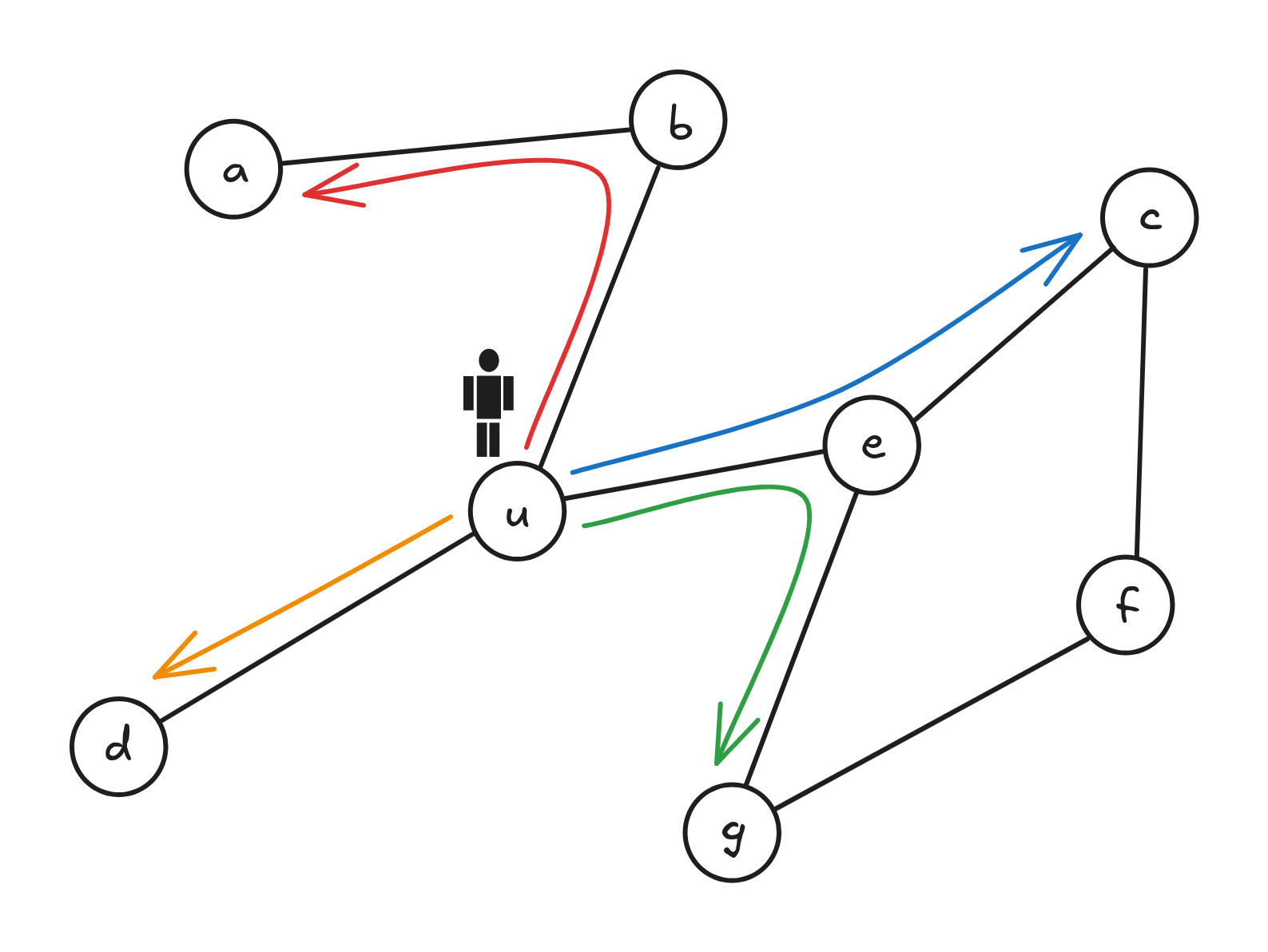}
      \caption{Walker on node $u$ with $d = 2$ has access to the information on the sequences of links and nodes highlighted by arrows, as they are the $d$ first nodes and links of the branches in $\mathcal{B}(u)$.}
    \label{fig:branches}
\end{center}
\end{figure}

On Figure~\ref{fig:branches}, we see the walker on node $u$ with $d = 2$ considers the sequence of nodes in red, green, yellow and blue, as they are the $d$ first nodes of the branches the walker can see. The walker can then evaluate the values of the two criteria on the nodes and links of the branch, and use them to decide its next move.

We notice that, in the set $\mathcal{B}(u)$, there can be multiple branches passing through each neighbor in $N(u)$. In a line of nodes, a single street, there is only one simple path in a given direction, and therefore the branch is unique.  However, at intersections there are multiple branches. For example, in Figure~\ref{fig:branches} we see the green and blue branches both start from node $u$ but go into different directions after an intersection at node $e$.

Another comment concerns the length of branches with respect to the vision depth. We see that, in the case of the orange arrow, a branch can be shorter than $d=2$, as a simple path can not go back to a node already visited. The walker on $u$ then only considers the nodes up to the end of the branch, i.e., $u$ and $d$.

\subsection{Weighting Branches}
\label{sec:rules}

We know that $n_u(t)$ and $J_{v\rightarrow u}(t)$ are two commensurable quantities. Indeed, the number of walkers is a quantity that is directly comparable to the net flux, as a group of walkers all moving from $u$ to $v$ at $t-1$ produces a net flux so that $J_{u\rightarrow v}(t) = n_v(t)$. However, if their respective vision depths are different, we cannot compare the two criteria directly. We have to define a way to aggregate them, taking into account a walker might see more values from one criterion.

To do so, for a walker $i$, we denote by $w_B(d_n,d_j,t)$ the weight, at time $t$, of a branch $B \in \mathcal{B}(x_i(t))$ 
and define it as:

\begin{equation}
    w_B(d_n,d_j,t) = \mathcal{N}_B(d_n,t)+ \mathcal{J}_B(d_j,t)
    \label{eq:branch}
\end{equation}

where the two terms are the mean of each criterion values the walker considers along the branch:

\begin{align}
    \mathcal{N}_B(d_n,t) &=  \sum_{i=1}^{\Delta_n} \frac{n_{b_i}(t)}{\Delta_n} \text{ \qquad  with } \Delta_n = \min{(|B|,d_n)}\\
    \mathcal{J}_B(d_j,t) &= \sum_{i=0}^{\Delta_j} \frac{J_{b_{i}\rightarrow b_{i+1}}(t)}{\Delta_j} \text{ \qquad    with } \Delta_j = \min{(|B|,d_j) -1}
\end{align}

This guarantees we sum comparable quantities, as each term in $\mathcal{N}_B(d_n,t)$ and $\mathcal{J}_B(d_j,t)$ is divided by the number of terms in its sum. We can finally define a walker tactic as follows.

\begin{definition}[Tactic]
    For a walker $i$ at time $t$, following the tactic defined by the pair $(d_n, d_j)$, is to identify the branch $B \in \mathcal{B}(x_i(t))$ with the maximum weight $w_B(d_n,d_j,t)$, and subsequently moves to the neighbour $v\in N(x_i(t))$ so that $v=b_1$.
\end{definition}
Notice that, in the case of several branches with the same highest value, the walker picks one node randomly among them.

\section{Experiments}
\label{sec:experiments}
In this section we measure how well our walkers perform at flocking. We seek to understand how the vision depth impacts their dynamics and if this new feature improves flocking behaviors. We run all experiments on the discretized Paris street network for 1000 steps with as many walkers as there are nodes in the network. They all start at random nodes.

To understand the impact of the vision depth on walkers behavior, we first measure the impact of the vision depth on the gathering score for each individual criterion. We will then allow walkers to make use of both criteria, with different vision depth for each of them, in order to identify the best combination.

\subsection{Impact of Vision on each Criterion}
\label{sec:eachvision}
\begin{figure}[h!]
\centering
    \includegraphics[width=\linewidth]{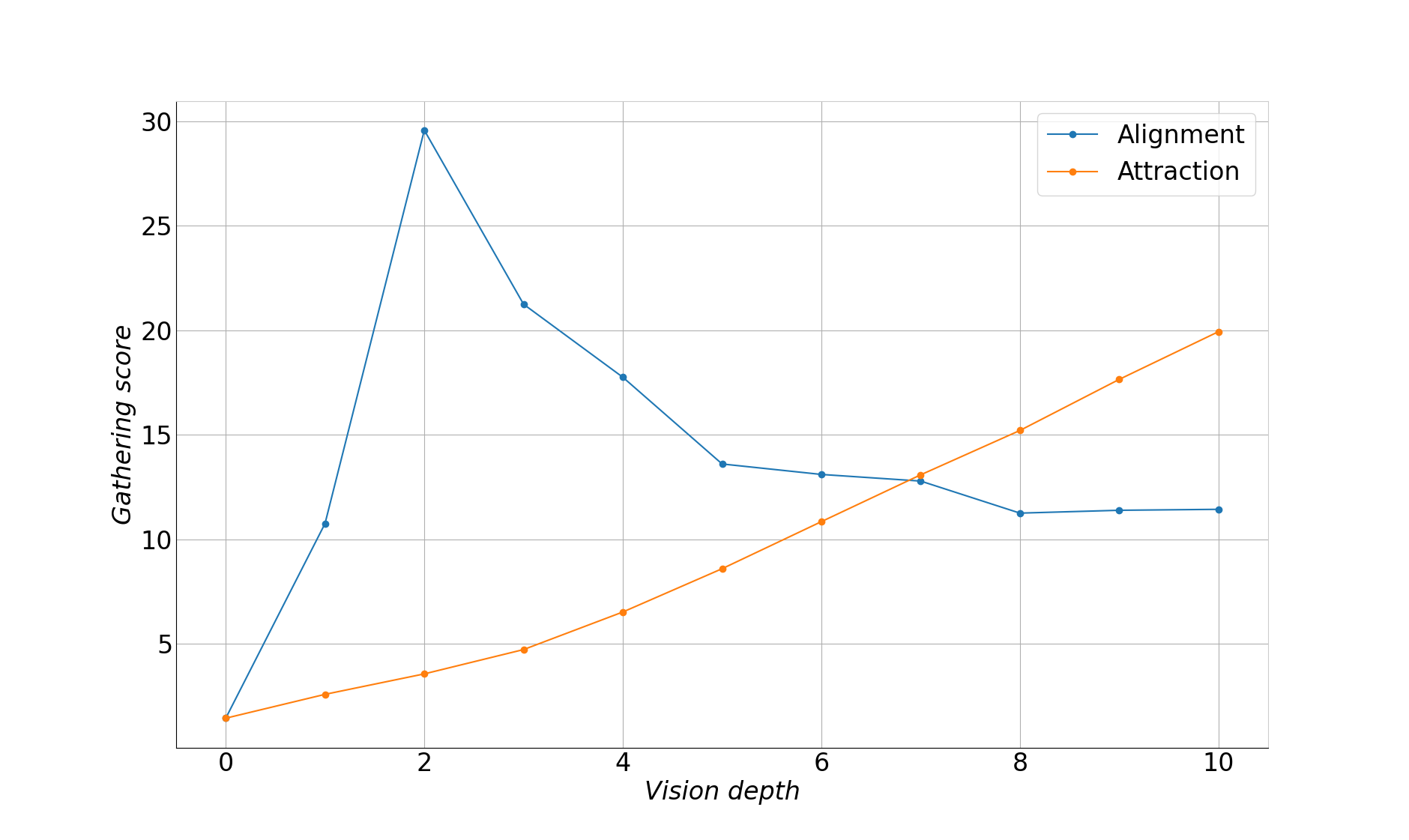}
    \caption{Evolution of the last step gathering score for the alignment and attraction rules with different vision depths.}
    \label{fig:vision}
\end{figure}

In this first experiment, walkers only make use of the attraction rule. In Figure~\ref{fig:vision}, we display the gathering score we obtain at the end of a run for each vision depth. It linearly increases with the vision depth. This is because walkers can see further and therefore interact with a number of walkers that is proportional to their vision depth. The gathering score is then roughly two times the vision depth, as a walker in a street sees exactly $2d_n$ nodes.

In the second experiment, walkers only make use of the net flux criterion. We know from~\cite{moinard2024fast} that this alignment rule, alone, creates a much richer process that does not simply converge after a few steps. That is why we see that, still on Figure~\ref{fig:vision}, this alignment rule is already better than the attraction one when walkers only see their neighbors ($d_n = d_j =1$).

Moreover, its gathering score still increases with the vision depth. In~\cite{moinard2024fast} authors showed that, each rule taken separately, the net flux criterion is the most effective at gathering walkers, except that it can not prevent a group of walkers from splitting into two groups at an intersection. However, with $d_j=2$, when a group of walkers splits at an intersection, the walkers that form the smallest new group now still perceive a positive net flux from the other walkers last move. They therefore come back and follow their original group. This means that, given $d_j=2$, the alignment rule is able to have all walkers in the network flock together at long time.

However, for higher values of the vision depth, the gathering score decreases. This is because a walker can now see too far and therefore anticipates the arrival of other walkers from their net flux. In such conditions, a small group will align with a bigger in advance; its walkers will turn back and be repealed by the bigger group before the two merge. This results in groups of walkers avoiding each other, and therefore not gathering despite exploring the network.

\subsection{Impact of Vision on Combinations of Criteria}
\begin{figure}[h!]
\centering
    \includegraphics[width=\linewidth]{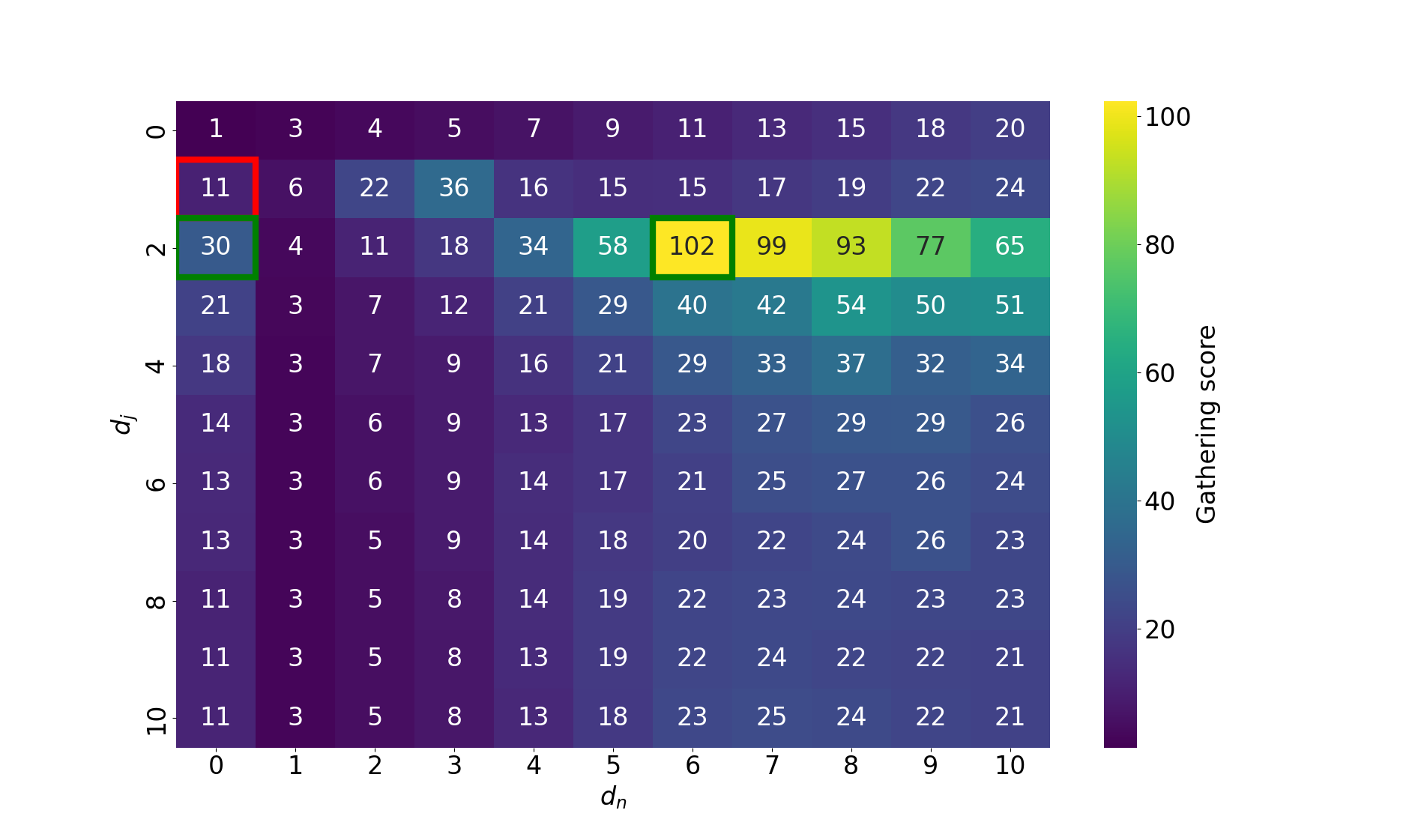}
    \caption{Heatmap of gathering scores for every tactic with combinations of vision depths $d_n$ and $d_j$ from $0$ to $10$.}
    \label{fig:heatmap}
\end{figure}

We run the same experiment as before,  with different vision depths for each criterion. We take only in count the gathering score at the last step of a run to see how different combinations of criteria play on flocking behavior. We display our results in Figure~\ref{fig:heatmap}.

Notice that the top most line and the left most column correspond to the experiments we presented in previous sections. With $d_n=d_j=0$, the top left cell is the result for random walkers.

First, the second column stands out from the previous and following columns, because of its poor scores. This is due to the vision depth for the attraction rule being equal to or smaller than the vision depth for the alignment rule. This implies that, while the alignment rule averages over multiple nodes, with some of them being empty, the attraction rule only takes into account the neighbor node. The contribution from the first term in equation~\ref{eq:branch} is then most of the time greater than the second term. Therefore, rather than slightly influencing a group that already flocks, the attraction rule becomes dominant. As this rule do not produce flocking on its own, the bigger the alignment vision depth, the more the score drops until being as bad as for the attraction rule alone.

This dynamic also explains why a value for at a cell $(x,y)$ in the upper right triangle of the matrix is better than the one at $(y,x)$ in the lower left one. Indeed, the upper right triangle corresponds to tactics for which the attraction vision depth for is greater than the alignment vision depth. In those cases, the attraction rule is able to influence the group that flocks, and the alignment rule is able to make the group flock. This is the best combination of vision depths for the two rules. This result is in line with the literature on flocking in continuous spaces~\cite{giardina2008collective}, where rules often do not apply in the same neighborhood. More precisely the attraction rule generally applies to a further neighborhood than the alignment rule.

Notice the first column and row do not respect such a pattern, as they do not describe combination of the two rules. Moreover, in the bottom right corner, a few cells do not either, although the value gap between such cells and their symmetric cell in the matrix is very small.

In Figure~\ref{fig:heatmap} we mark with a red square the cell showing the gathering score for walkers using the alignment rule with $d_j=1$, like in~\cite{moinard2024fast}. We also highlight in green squares the two main contributions of this paper.

We first have the alignment rule with $d_j=2$, that greatly increases groups size as we explained in Section~\ref{sec:eachvision}. We now also see an optimal combination of the two rules, with vision depths $d_n=6$ and $d_j=2$. This combination takes advantage of the flocking behavior the alignment rules induces, and of the gathering effect from the attraction rule. This gives rise to groups that flock while being to move preferentially towards other groups. Such groups are up to 10 times larger than previous results. Increasing $d_j$ obviously degrades the gathering score due to the repealing effect we highlighted. However, although it is not clear why, increasing $d_n$ above $6$ also worsen results. We also notice that, the further down the y-axis, the more the optimal combination on each line shifts to the right.  

\begin{figure}[h!]
\centering
    \includegraphics[width=\linewidth]{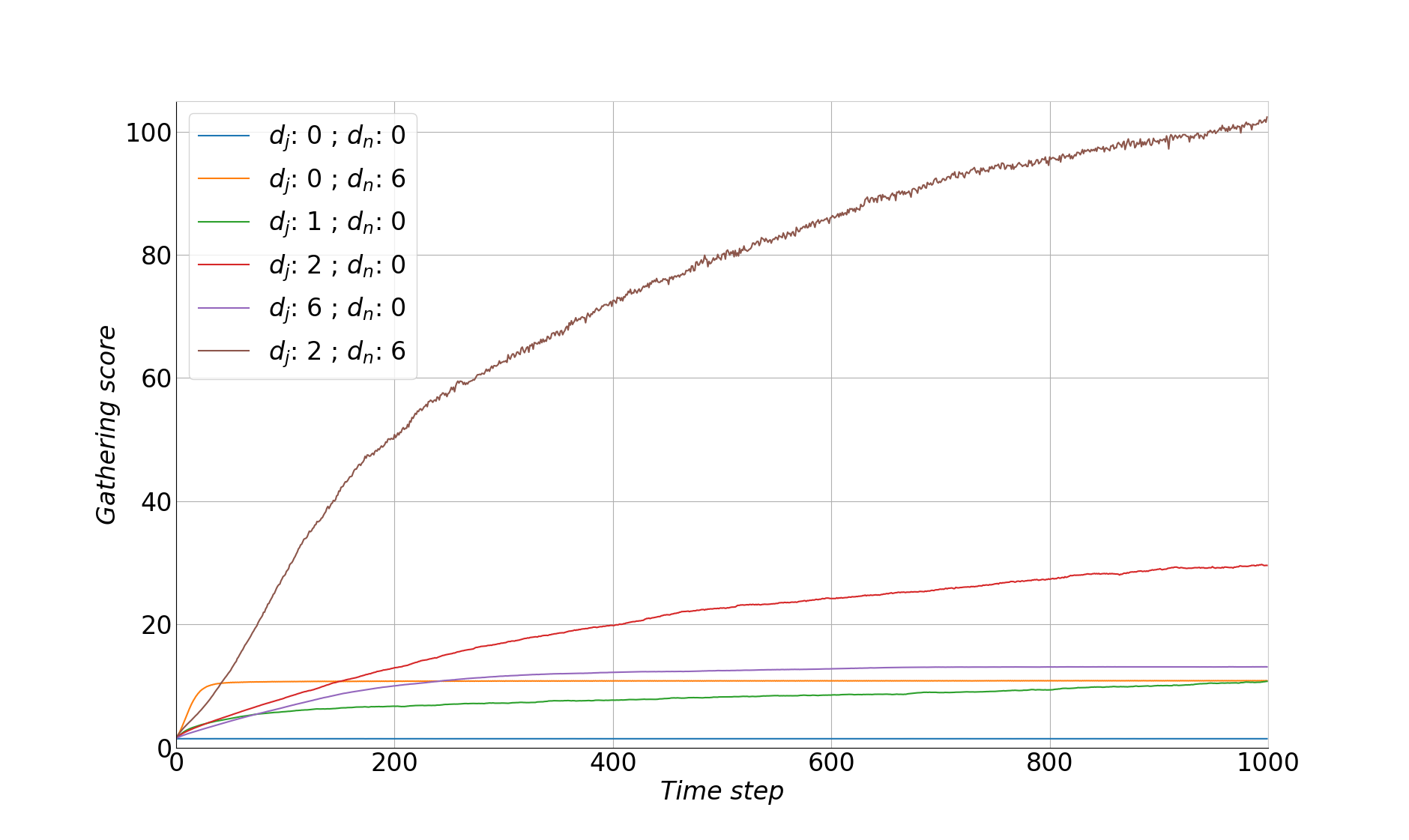}
    \caption{Evolution of gathering score for some relevant tactics with different vision depths.}
    \label{fig:rez_vision}
\end{figure}

Finally, we display the evolution of the most interesting tactics in Figure~\ref{fig:rez_vision}. All tactics using the attraction rule alone, whatever the value of $d_n$, are easy to describe; groups quickly form and do not evolve anymore, walkers being unable to move towards other groups that are now out of their range of vision. We notice that, in the first time steps and for big $d_n$ value, walkers using attraction alone can form larger groups than those using any combination of the two rules. Therefore, if walkers only have a very limited amount of time to gather, the attraction rule alone is the best choice. However, if walkers have more time to gather, the combination of the two rules clearly becomes the best option. 

\subsection{Robustness of Groups}
\label{sec:robustness}

It is crucial for protesters to form groups that resist adversary forces that may break them up. In order to explore this robustness, we follow the method of~\cite{moinard2024fast} that models break ups as walkers suddenly following a random walk for one step. In this way, a group located at a given node splits into smaller groups that move to neighbor nodes, in a way similar to a group of protesters targeted by adversary forces.

More formally, we perform the following experiment: we run a simulation for $500$ steps, then we impose walkers to move at random for one step. Finally, walkers use once again their combination of rules until the end of the run. We performed this experiment with the two tactics circled by the green squares in Figure~\ref{fig:heatmap}, i.e., the best tactic we found and the alignment rule with $d_j=2$, and the tactic circled in red with $d_j=1$.

\begin{figure}[h!]
\centering
    \includegraphics[width=\linewidth]{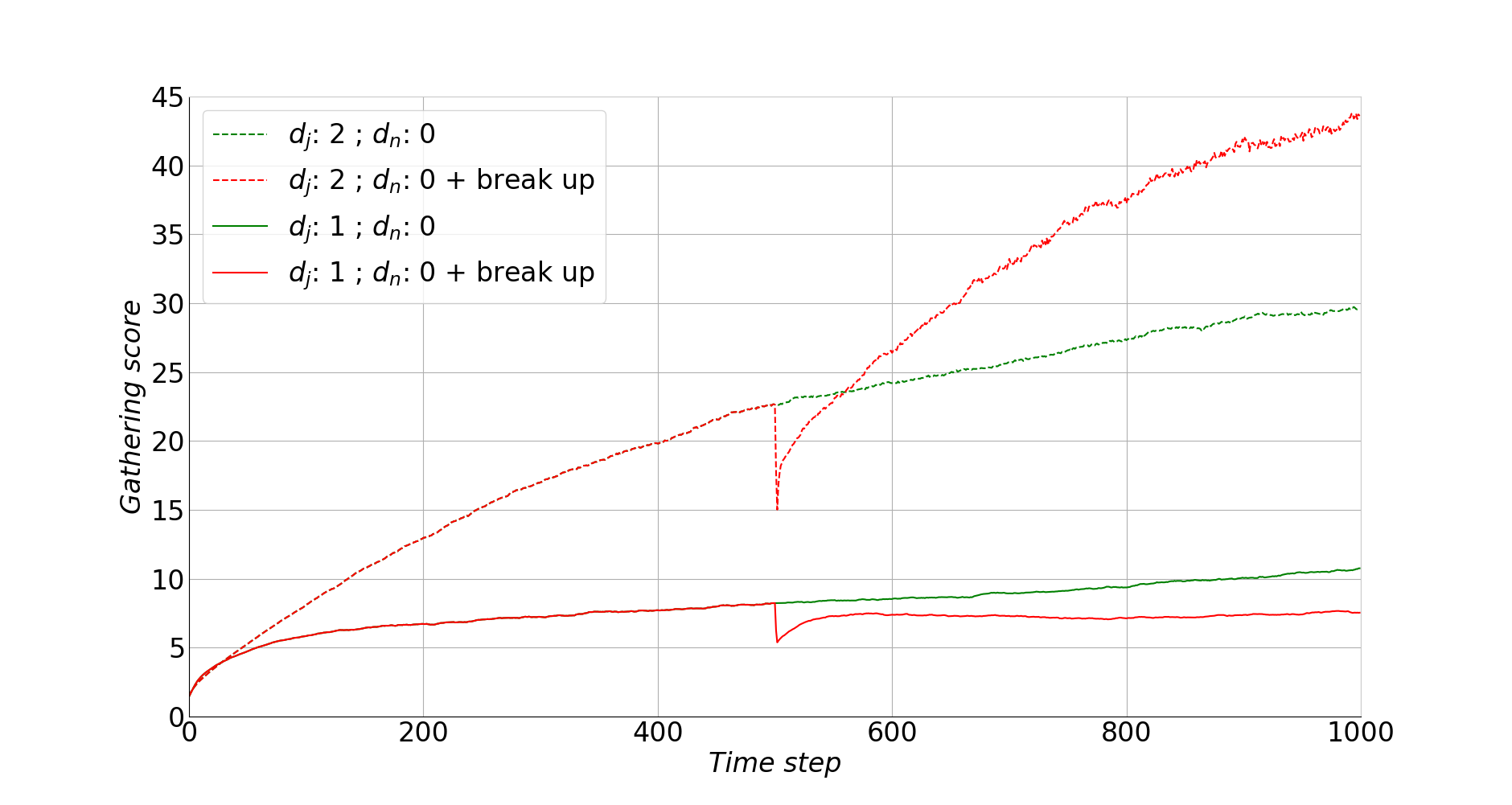}
    \caption{Plot of robustness experiments, similar to Figure~\ref{fig:rez_vision}. We display the gathering scores for tactics without break ups in green, with a break up in red.}
    \label{fig:rob}
\end{figure}

Figure~\ref{fig:rob} displays the observed scores 
for the two latter, as the combination gave similar result to the alignment with $d_j=2$, indicating that the attraction rule is not a key factor in the robustness of groups.

At the break up, groups size drops drastically. Then, for the alignment rule with $d_j=1$, new groups are smaller than they would have been without the break up. This is a result already established in~\cite{moinard2024fast}. However, walkers following a tactic with $d_j=2$ are able to recover from the break up and form groups even bigger than before. This is thanks to the longer vision depth that allows the walkers to interact further and therefore to recover from the break up.

In~\cite{moinard2024fast}, combining multiple rules was enough to create robust groups that would recover from break ups. The gathering score would go back to its initial value as if the break up never happened. However, we show here that the vision depth is also a key factor in the robustness of the groups that enables group to exhibit a form of anti-fragility, i.e., turning bigger after a break up.

\section{Conclusion}
\label{sec:conclusion}
In this paper, we  presented a novel approach to produce flocking behaviors on a network which includes a parameter for the vision range of walkers. We studied our model on street networks, as they are a very relevant application case. 

Doing so, we obtained better results than the original model in~\cite{moinard2024fast}. Indeed, expanding the vision of walkers for the alignment rule guarantees walkers do not split into multiple groups at intersection. Moreover, a tactic that combines attraction and alignment rules, with different vision depth for each, leads us to results ten times greater and such groups display remarkable robustness.

This simple model paves the way for studying other characteristics specific to flocking on a large variety of networks. Moreover, it can help to model and understand the behavior of pedestrians in urban environments.\\

\paragraph{Reproducibility.}

We provide an implementation of our models in C, and a Python code result analysis, with documentation, at: \url{https://gitlab.com/guillaume_moinard/public-vision-flock}

\paragraph{Acknowledgements.}

This work is funded in part by the CNRS MITI interdisciplinary programs.

\end{document}